\documentstyle[graphicx,longtable]{aipproc}

\newcommand{\CP}{\mbox{\em CP}}
\newcommand{\T}{\mbox{\em T}}

\begin{document}
\begin{flushleft}
\small
Report to the 14th International Symposium on Spin Physics \\
October 16--21, 2000, RCNP, Osaka University, Osaka, Japan
\end{flushleft}
\title{Physics Beyond SM at RHIC with Polarized Protons}

\author{A.~Ogawa$^{\dag}$, V.~L.~Rykov$^{\S}$\thanks{Supported in part
    by the U.S. Department of Energy Grant DE-FG0292ER40713.} and
  N.~Saito$^{\ddag}$\thanks{This work has been done partly within the
    framework of RIKEN RHIC-Spin project.}}
\address{$^{\dag}$Penn State University, University Park, PA 16802, USA\\
$^{\S}$Wayne State University, Detroit, MI
48202, USA\\
$^{\ddag}$RIKEN--BNL Research Center, Upton, NY 11973, USA}

\maketitle

\begin{abstract}
The capabilities of RHIC with polarized protons to test the Lorentz
structure of electroweak interactions and also the properties of MSSM
Higgs, should it be discovered, are discussed.
\end{abstract}

\section*{Introduction}

RHIC-Spin experiment is about being started. It has the solid
program\footnote{See plenary talk by N.~Saito in this proceedings.},
mainly concentrated around the measurements of proton spin dependent
structure functions. For the last years, few attempts have been
undertaken to also evaluate RHIC-Spin capabilities for testing physics
beyond the Standard Model
(SM)~\cite{virey:95:97,rykov:98:99,soffer:00}. Since the recent
suggestion~\cite{roser:00} on a considerable increase of RHIC
luminosity in $pp$\, mode along with a sizeable energy increase,
RHIC-Spin potential in this area became looking even more
promising. 

In this report, we update our earlier study~\cite{rykov:98:99} of
RHIC-Spin capabilities to explore the Lorentz structure of electroweak
interactions and also provide some estimates on whether Higgs sector
could be reachable at RHIC. In our considerations, we assume
\mbox{$\sqrt{S} =$ 500 GeV} and the luminosity of 0.8~fb$^{-1}$/year
for RHIC before an upgrade (RHIC-500), and \mbox{$\sqrt{S} =$ 650
  GeV} and the luminosity of $\sim$20~fb$^{-1}$/year for the after
upgrade (RHIC-650).

\section*{Lorentz structure of quark electroweak current}

Lorentz structure of electroweak interactions had always been and
remains to be the focus of many precise nuclear and particle physics
experiments. Over the last few years, the first measurements of the
proton's weak magnetism have been accomplished~\cite{mueller:97}.
Much attention had also been paid to test the universality of the
electroweak interactions, in general, and to evaluating electroweak
dipole moments of quarks and leptons, in particular. The current
experimental constraints to magnetic and electric dipole moments of
quarks and leptons, except electrons and muons, are still quite far
from the values, predicted in the SM. If any of these moments were
found to be nonzero and above SM's expectations, this would be a clear
signal of a new physics beyond SM~\cite{escribano:94,escribano:93:97}.

The most stringent experimental constraints, applicable to all
components of quark and $\tau$-lepton dipole moments, come from the
recent analyses~\cite{escribano:94,escribano:93:97} of electroweak
data from high energy colliders. In these analyses, it has been
assumed that theories beyond the SM, emerging at some characteristic
energy scale above $W$/$Z$\, mass, have effects at low energies
\mbox{$E\leq M_{W,Z}$}, and these effects can be taken into account by
considering a Lagrangian that extends the SM Lagrangian, $L_{SM}$:
$L=L_{SM}+L_{eff}$. To preserve the consistency of the low energy
theory, it had also been assumed that $L_{eff}$\, is
SU(3)$\bigotimes$SU(2)$\bigotimes$U(1) gauge
invariant. Phenomenologically, this extension is equivalent to an
introduction of a tensor coupling of fermions to gauge bosons which
had also been discussed in Refs.~\cite{kane:92,rykov:98:99,soffer:00}:
\begin{equation}
L_{eff}^{charged} = \frac{g}{2\sqrt{2}\cdot\Lambda}\Big\{
\overline{q}_{d}\sigma^{\mu\nu}(f_{T}^{+}+f_{T}^{-}\gamma_{5})q_{u}
\partial_{\nu}W_{\mu}^{-} +
\overline{q}_{u}\sigma^{\mu\nu}
(f_{T}^{^{*}\!\!+}-f_{T}^{^{*}\!\!-}\gamma_{5})q_{d}
\partial_{\nu}W_{\mu}^{+}\Big\}
\label{eq:lagrangian}
\end{equation}
In Eq.~(\ref{eq:lagrangian}), representing the charged-current
part\footnote{With omitted quadratic terms, proportional to
  $g^2W_{\mu}W_{\nu}$; the neutral-current $L_{eff}$\, looks
  similarly.} of $L_{eff}$, $g$\, is the electroweak coupling
constant, $\Lambda$\, is the energy scale of the ``full strength''
tensor interactions, and the asterisk denotes the complex
conjugate. The notations $q_{u}$\, and $q_{d}$\, are for the ``upper''
\mbox{($u$\, and $c$)}\footnote{$t$-quark is virtually not reachable
  at RHIC.} and ``lower'' ($d, s, b$) quarks, respectively. The \CP-
and \T-invariance of model (\ref{eq:lagrangian}) is broken if any or
all ``formfactors'' $f_{T}^{\pm}$\, were complex\footnote{In this
  paper: $\gamma_{5}= -i\gamma^{0}\gamma^{1}\gamma^{2}\gamma^{3}$\,
  and $\sigma^{\mu\nu} = \frac{1}{2}(\gamma^{\mu}\gamma^{\nu} -
  \gamma^{\nu}\gamma^{\mu})$.}.

With the coupling~(\ref{eq:lagrangian}), a number of prohibited in the
SM spin asymmetries must show up in the annihilation of polarized
$q\overline{q}\rightarrow W^{\pm}/Z^{0}/\gamma\rightarrow
l\overline{l}$~\cite{rykov:98:99,soffer:00}. For polarized proton
collisions, particularly interesting would be single-spin asymmetries
with transverse polarization, arising from the interference of SM's
and tensor couplings, because: a) these asymmetries are strongly
suppressed in SM and b) they are expected to have a good sensitivity
to anomalous interactions due to quite strong correlations between the
proton spin and polarizations of high-$x$\, valence quarks, that
participated in gauge boson production~\cite{soffer:98}. The
triple-vector correlations, \mbox{$\propto$\boldmath
  ($k\cdot$[$\zeta^{\perp}_{q}\times p_{q}$])}, give rise to the
``left-right'' asymmetry $A_{N}$, while nonzero ``up-down'' asymmetry
$A_{T}$\, comes from {\em P}- and \CP-violating\footnote{In this
  particular model for $L_{eff}^{charged}$.} two-vector correlations,
\mbox{$\propto$\boldmath ($\zeta^{\perp}_{q}\cdot k$)}. In the
formulae above, \mbox{\boldmath $p_{q}$}\, and \mbox{\boldmath
  $\zeta_{q}^{\perp}$}\, are for the momentum and transverse
polarization of an incident quark, and \mbox{\boldmath $k$}\, is the
momentum of a final lepton. The respective asymmetries
$\hat{a}_{N,T}$\, in $W^\pm$\, production at the quark interaction
level, integrated over the phase space of final leptons, would be as
follows:
\small
\begin{eqnarray}
\hat{a}_{N} & \simeq &
\frac{3\pi}{16}\cdot\frac{M_{W}}{\Lambda}\cdot
\mathop{\mathrm{Re}}\{f^{+}_{T}\mp f^{-}_{T}\}\Big/
\Big\{1 + \frac{M_{W}^2}{4\Lambda^2}(\mid f_{T}^{+}\mid^2+\mid
f_{T}^{-}\mid^2)\Big\}
\label{eq:an} \\
\hat{a}_{T} & \simeq &
\frac{3\pi}{16}\cdot\frac{M_{W}}{\Lambda}\cdot
\mathop{\mathrm{Im}}\{f^{-}_{T}\mp f^{+}_{T}\}\Big/
\Big\{1 + \frac{M_{W}^2}{4\Lambda^2}(\mid f_{T}^{+}\mid^2+\mid
f_{T}^{-}\mid^2)\Big\}
\label{eq:at}
\end{eqnarray}
\normalsize
Formulae (\ref{eq:an},\ref{eq:at}) have been obtained, using the
solutions of Ref.~\cite{rykov:98:99} and with the assumptions that,
the SM part of $q\overline{q}W$-interactions as well as the lepton
coupling to $W^\pm$\, were purely $V$-$A$. The upper and lower signs
correspond to $W^{+}$\, and $W^{-}$\, productions, respectively.
\begin{table}[htb]
\caption[]{RHIC-Spin sensitivity to $q\overline{q}W$\, tensor
  coupling. Constraints (2$\sigma$) in columns 2--3 are for the
  asymmetries at quark ($\hat{a}$) and proton
  ($A\simeq\hat{a}/2$~\cite{soffer:98,soffer:00}) interaction
  levels. RHIC sensitivities (2$\sigma$) to $A_{N,T}$ are shown in
  columns 4--5, assuming a 70\% proton beam polarization.}
\label{tab:moments}
\begin{tabular}{||c|c|c||c|c||} \hline
\multicolumn{3}{||c||}{Constraints from
  Ref.~\cite{escribano:94}} & \multicolumn{2}{c||}{RHIC sensitivity in
  ($W^+/W^-$) modes} \\ \hline 
\mbox{\footnotesize$(\mid f_{T}^{+}\mid^2 + \mid f_{T}^{-}\mid^2)\cdot
M_{W}^2/\Lambda^2$} & $\hat{a}_{N,T}$ &
$A_{N,T}$ & RHIC-500 & RHIC-650 \\ \hline
$\leq$0.15 & $\lesssim$(20--30)\% & $\lesssim$(10--15)\% &
$\;\;\;\;\;\;\;\sim$(1.5/3.0)$\%\;\;\;\;\;\;\;$ & $\sim$(0.2/0.4)\% \\
\hline
\end{tabular}
\end{table}

The experimental constraint on ``formfactors'' $f_{T}^{\pm}$\, in
charged current $u\leftrightarrow d$\, transitions, converted to the
convenient for our goals representation, is shown in
Table~\ref{tab:moments}, along with the respective limits on spin
asymmetries in $pp^{\uparrow}\rightarrow W^{\pm}+X\rightarrow
l\overline{l}+X$\, processes. These limits then are compared with the
RHIC sensitivities to $A_{N,T}$. One can observe that the current
limits on $f_{T}$\, in quark sector could be lowered by a factor of
5-10 at RHIC-500, and by almost an order of magnitude more after the
upgrade to RHIC-650. And what is probably even more important, the
real and imaginary parts of $f_{T}^{\pm}$\, could be constrained
individually from four independent measurements of $A_{N}$\, and
$A_{T}$\, in $W^+$\, and $W^-$\, productions.

\section*{Will Higgs physics be reachable at RHIC?}

In the recent years, extensive efforts for cornering Higgs bosons
resulted in the significantly shrunken area of the still allowed Higgs
sector parameters. The actual discovery of Higgs particle(s) is
expected to occur in not so remote future. Then, the focus will be
shifted to studying their properties. Does RHIC with {\em polarized\/}
protons have capabilities to contribute in this study?

The production cross section of Higgs strongly depends on the
helicities of initial gluons and quarks. As a result, measurements
of spin-correlations in polarized $pp$\, collisions may allow us,
for example, to determine \CP-parities of Higgs states. Potentially,
an interference between Higgs production and other SM processes in
collisions of polarized gluons and quarks can generate a number of
interesting single- and double-spin asymmetries\footnote{Similar to
  the ones described, for example, in Refs.~\cite{akawa:00}.},
including those sensitive to \CP-violation in Higgs sector. In some
cases, the polarization may help to improve signal-to-background ratio
compared to unpolarized particle collisions.

Currently, the data favor low mass Higgs, probably just around
$\sim$100~GeV, and for the Minimal Supersymmetric extension of the
Standard Model (MSSM), the versions with large $\tan\beta\gtrsim$~10
seems as taking preferences\footnote{See, for
  example,~\cite{pdg:00,carena:00} and references therein.}. Let's
imagine for a moment that, just before RHIC-Spin upgrade, a neutral
scalar\footnote{Or scalars; for large $\tan\beta$, the theory allows
  for all three MSSM neutral bosons, $h^0$, $H^0$\, and $A^0$, to be
  sitting simultaneously in the mass region from $\sim$100 to 130
  GeV~\cite{carena:00}.} has been discovered with the mass
$M_h\sim$115-120~GeV, and its characteristics were consistent with the
MSSM Higgs for $\tan\beta\simeq$~30. The estimated production cross
section, $\sigma_h$, for such boson(s) at RHIC-650 would be
$\sim$0.5~pb with approximately equal contributions of $gg\rightarrow
h$\, and $q\overline{q}\rightarrow h$\, subprocesses\footnote{Notation
  $h$\, is used here as a generic name for either $h^0$, $H^0$\, or
  $A^0$.}. For an integrated luminosity of 20~fb/year, one may expect
a yield of $\sim$10$^4$~$h$/year.

The main decay modes of low mass Higgs scalars are: $h\rightarrow
b\overline{b}$ with the branching of $\sim$90\%, and
$h\rightarrow\tau^+\tau^-$\, with the branching of
$\sim$8-9\%. At the early stage, both these modes have been determined
as hardly be suitable for the Higgs searches. However, as it was
pointed out in Ref.~\cite{gunion:90:1}, $h\rightarrow\tau^+\tau^-$\,
decay might become useful ``to provide confirming evidence for a
signal found in other modes''\footnote{See   also
  Ref.~\cite{carena:00}, page 29 for the similar qualification.}. In
this report, as a first step toward understanding the Higgs-at-RHIC
problem, we will start with the evaluation of only these two modes
with the highest branchings, keeping in mind of course, that future
studies could reveal better ways for the Higgs sector exploration at
RHIC\footnote{See, for example, J.~F.~Gunion and T.~C.~Yan, {\it
    Phys. Rev. Lett.,} {\bf 71} (1993) 488.}.


The main obstacle for using $b\overline{b}$\, channel at hadron
colliders is the large QCD background from $gg\rightarrow
b\overline{b}$. At RHIC-650, the estimated with PYTHIA production
cross section $\sigma_{QCD}^{b\overline{b}}$ for $b\overline{b}$-pairs
in the mass interval of 115$\times$(1$\pm$10\%)~GeV is
$\gtrsim$10$^3$~pb. Resulting from cross section estimates Higgs
signal-to-background ratio, 
$\sigma_h/\sigma_{QCD}$, at $\lesssim$5$\cdot$10$^{-4}$\, does not
seem too encouraging,
particularly for studying Higgs properties. Potentially, an
interference between $gg\stackrel{QCD}{\rightarrow}b\overline{b}$\,
and $gg\stackrel{h}{\rightarrow}b\overline{b}$\, could improve this
ratio up to $\sim\sqrt{\sigma_h/\sigma_{QCD}}\sim$~10$^{-2}$.
Unfortunately,
in the ultra-relativistic limit of massless $b$-quark, these two
channels do not interfere. As a result, the actual interference term
is additionally suppressed by a factor $m_b/M_h\sim$~0.03-0.04, which
brings it back to the same low level of $\lesssim$5$\cdot$10$^{-4}$.


The background to $\tau^+\tau^-$\, pairs from Higgs decays
predominantly comes from the Drell-Yan process:
$q\overline{q}\rightarrow\gamma/Z^0\rightarrow\tau^+\tau^-$. For
RHIC-650, PYTHIA estimates the Drell-Yan cross section for
$\tau^+\tau^-$\, at $\sim$0.25~pb in the $\tau$-pair mass interval of
115$\times$(1$\pm$10\%)~GeV. Taking into account $\sim$8\% branching
of Higgs to $\tau^+\tau^-$, the signal-to-background ratio in this
mode would be $\sim$15\%. With the detection efficiency for high-mass
$\tau$-pairs at about 20\%, the event rate in the mass interval above
should be expected at $\sim$10$^{3}$~$\tau$-pairs/year. For this rate,
$\sim$15\% excess of events due to Higgs decays should be detectable
well above the statistical fluctuations.
Then,
for example the \CP-parity of a detected Higgs boson could
be determined by measuring the sign of the contributing to the cross
section double-spin correlation \mbox{\boldmath
  ($\zeta^{\perp}_{q}\cdot\zeta^{\perp}_{\overline{q}}$)} in
collisions of transversely polarized protons.
However, the statistics of just $\sim$10$^2$\, of
$h\rightarrow\tau^+\tau^-$\, decays/year may not be sufficient for
measuring double-spin asymmetries. Unfortunately, we cannot count on
potentially more sensitive single-spin ones which could arise from an
interference of two competing $q\overline{q}\rightarrow\tau^+\tau^-$\,
channels. This interference will be vanishingly small because
Drell-Yan pairs are mainly produced by light quarks, while the
respective part of Higgs cross section will predominantly be due to
$b\overline{b}$-annihilation.

\section*{Conclusion}

It has been shown that RHIC with polarized protons will have highly
competitive capabilities for hunting anomalies in the Lorentz
structure of electroweak interaction due to physics beyond SM.

The low-mass Higgs might be reachable at high luminosity RHIC-650,
although finding appropriate modes to study polarization phenomena in
Higgs sector will be a quite challenging task.

Authors appreciate the valuable discussions with D.~Boer,
D.~A.~Cinabro, S.~Dawson, R.~L.~Jaffe, J.~S.~Lange, T.~Maruyama and
J.~Soffer. One of us (VLR) is thankful to T.~M.~Cormier for the
support of his participation in SPIN2000 Symposium.

\end{document}